**Title:** Analytical solution of the Schrödinger equation for the neutral helium atom in the ground state considering the uncertainty principle, vibrational modes and quantum-electrodynamical effects

**Author:** Frank Kowol, PhD., Germany

**Email:** frank.kowol@gmx.net

**ORCID-ID:** 0009-0000-4189-9372

# *Analytical solution of the Schrödinger equation for the neutral helium atom in the ground state considering the uncertainty principle, vibrational modes and quantum-electrodynamical effects*

Frank Kowol, Ph.D., Germany (frank.kowol@gmx.net)

## 1. Abstract

This report presents a direct ab initio solution of the Schrödinger equation for neutral helium or helium-like systems and reproduces its energy level of the ground S state $^1S_0$. Redefining the wave function with tools of complex analysis, the entangled two-electron system considering *S=L=0* is analytically derived from the Schrodinger equation by using Laplace transformations and examined in detail.

Though treating electrons point like, yet we understand their electrical field not caused by a point charge, but respecting Heisenberg's uncertainty principle we solve the Schrödinger equation with an adapted potential independent of known Hartree approaches. Hence, a generic electron potential is derived and integrated into the equations. We propose a significant energy contribution considering dark electronic vibrational modes, and by utilizing perturbation calculations, we take contributions from the Breit Hamiltonian as well as Quantum-electrodynamical effects like the Lamb Shift into account. By this approach the ground state energy could be reproduced to an accuracy of +1.86 meV referred to the literature value.

With this model and including the vibrational electronic modes we can contribute to a proposed explanation for the inertness of helium regarding chemical reactions, i.e., the principle of the closed shell, as well as the spatial structure of helium, giving reasonable results compared to literature values, thus providing us with a powerful model to describe helium analytically.

## 2. Motivation and approach

The neutral helium atom is one of the trickiest problems in modern physics, as it combines the apparently well-understood structure of a quantum-mechanical hydrogen-like system and well-defined quantum numbers with the complex three-body problem. Since the work of Hylleraas [4], an extreme variety of attempts have been made to solve the Schrödinger equation, including numerical approaches and variation calculus which yield great success over the years [3-23]. A full historical breakdown would extend the scope of this article by far, but some milestones and their main actors should be named:

- Hylleraas' variational method (1929): Introducing the interelectronic distance $r_{12}$ in a trial wavefunction reduced the ground-state energy error to ≈0.01 eV with only a few terms [4,74]
- 1/Z perturbation expansion: Treating 1/Z as a small parameter converges only up to a critical value ≈1.09766, yielding only a handful of correct decimal places before divergence sets in [75]
- Pekeris (1958): First large-scale WEIZAC calculation using perimetric coordinates reached nonrelativistic energies accurate to 0.01 $cm^{-1}$ [76]
- Korobov's ECG basis (1999): Explicitly correlated Gaussians $exp(-\alpha\, r_1, \beta\, r_2, \gamma\, r_{12})$ achieved 25 decimal-place accuracy with N≈5200 terms [87]
- Schwartz (2006): A mixed basis of negative powers and a logarithmic term in $s = r_1 + r_2$ extended the border to N=10257 terms, pushing accuracy beyond 35 decimal places [2,3]
- Exponential Hylleraas-CI (2012): Combining linear $r_{12}$ factors with Slater-type exponentials further accelerating convergence and reporting nonrelativistic energies to 18 significant figures with $\leq 10^4$ terms [78]
- Mass-polarization & Relativistic Corrections: Subsequent numerical evaluations include mass-polarization and leading relativistic terms, enabling sub-MHz agreement with experiment. [79,80]

However, a non-analytical approach has the disadvantage that the basic mechanisms of action that ultimately lead to the binding energy remain opaque. A real understanding of the helium atom, its energy configuration and the mechanism of a closed shell stays in the dark. An analytical solution is therefore to be preferred in any case, but so far it has not been possible to realize this despite many attempts with various approaches and different wave function solutions [2,3,5,6,8,9,10,15,16,18,19,20]. The challenge is mainly caused by a three-body problem of a nucleon-electron-electron configuration with entangled particles that results in a non-linear Schrödinger equation, and as such cannot be solved trivially without further ado. The idea of a quasi-planetary system with the nucleus e.g. as a star and the electrons acting as planets is just as misleading as the idea of a sum of arbitrary wave function, especially since the system in the ground state of a $^1S_0$ -helium configuration has neither an orbital angular momentum nor a net total electron spin. It is easy to see that a multipole moment would always arise in the particle picture with two separately localized electrons and a nucleus, especially if one demanded a configuration of minimum energy with the electrons occupying quasi-opposite positions. Even if we assume, that a net dipole or multipole moment equals zero over time, there still emerges an angular momentum by rotation, which would be detectable. In the experiments such effects, regardless of the type, have never been observed [1,2 7,16] and are theoretically prohibited by quantum mechanics. To derive a valid analytical solution, a few steps are required, which are discussed in the following sections:

- The nature of the ground state for the electrons is examined in detail. Given the premise that both electrons must be described as one entangled wavefunction in a state with lowest possible energy we chose the approach describing the wave function by complex analysis calculus.
- The electrical field of electrons is described considering the uncertainty principle to access a general formulation for the potential. The decisive difference between hydrogen and helium - apart from the obvious fact of the double nuclear charge - is the presence of two entangled electrons, significantly altering their Hamiltonian.
- With the preliminary work, an analytical solution of the Schrödinger equation is presented utilizing Laplace transformation, followed by a comparison of the helium wave function with hydrogen as well as the Hylleraas wave function.
- To achieve the desired accuracy for the ground state energy, several perturbations are discussed including contributions by dark vibrational modes of the collective electron system as well as contributions from Breit terms, and nonetheless the Lamb shift is considered. Especially the concept of a dark vibrational electron mode opens a fascinating way to explain the closed shell effect.
- Finally, the analysis of the energy in the ground state $^1S_0$ is performed and compared with literature values. Summary and outlook as well as the appendices, the bibliography and the documentation of the MATLAB code finally round off this report.

## 3. The nature of the electronic ground state

Despite its alleged simplicity, the basic state of the helium atom already offers immense complexity. For calculations, the nucleus can be easily separated from the electron system due to the large difference in mass and the significantly smaller spatial dimensions in relation to the atomic radius; for the electrons the way is not so straight forward.

Before the ground state of helium can be investigated further, it is important to clarify a few important premises. [3,7,11,12,14,17,21]

- Both electrons are indistinguishable in principle, so the wave functions must be anti-symmetrized as the Pauli principle states. When considering the ground state, this leads to a $1s^2$singlet with anti-parallel electron spin resulting in a symmetry condition $|\psi,\phi\rangle = |\phi,\psi\rangle$ with $S = 0$.
- By definition, the ground state is the state with the lowest energy, so considering minimal energy contribution for the angular momentum in the radial Schrödinger equation $\frac{\hbar^2 l(l+1)}{2m_e r^2}$, this necessarily results in $l_i = 0$ for both electrons. The literature confirms this via the vanishing magnetic moment in the ground state. [15,17]
- At the same time, helium in the ground state - as known from the literature - does also not have an electric dipole moment $\langle\vec{p}\rangle$, so with $\langle\vec{r}_1\rangle, \langle\vec{r}_2\rangle$ defined as expectation values of the two electrons the following is valid:

$$\langle\vec{p}\rangle = q\langle d\rangle\vec{e}_{12} = 2e(\langle\vec{r}_1\rangle - \langle\vec{r}_2\rangle) = 0 \Leftrightarrow \langle\vec{r}_1\rangle = \langle\vec{r}_2\rangle \tag{1}$$

- Due to the central field of the core with $l_i = 0$ and $\langle\vec{p}\rangle = 0$ this consequently leads to the spherical symmetry of the state function. Consequently, we can limit ourselves to specifying the radius to the core $\vec{r}_i \longrightarrow r_i > 0$ for both electrons.

Obviously, both fermions share the narrow space around the nucleus and due to their entanglement they cannot be considered separately as coherent states but are a common system. A product approach according to $\alpha|\phi\rangle|\psi\rangle + \beta|\psi\rangle|\phi\rangle$ can certainly be a valuable solution approach for separated excited states, but for entangled ones such as ${}^1S_0$ it cannot represent the full solution space. Only an element from the tensor product space $|\psi,\phi\rangle = |\phi,\psi\rangle \in |\phi\rangle\otimes|\psi\rangle$ can do this.[14]

This of course also implies that these state functions must have the same energy as can be easily seen, if the full Hamiltonian for the Helium atom is written down:

$$0 = \Delta E = E_1 - E_2 = \langle\psi,\phi|\hat{H}|\psi,\phi\rangle - \langle\phi,\psi|\hat{H}|\phi,\psi\rangle \tag{2}$$

With the well-known formula [15,18] for $\hat{H} = \hat{H}_1 + \hat{H}_2 + \hat{H}_{ee}$ defined by:

$$\hat{H}_1 = \frac{\hat{p}_1^{\,2}}{2m_e} - \frac{Ze^2}{4\pi\varepsilon_0|r_1|} \quad and \quad \hat{H}_2 = \frac{\hat{p}_2^{\,2}}{2m_e} - \frac{Ze^2}{4\pi\varepsilon_0|r_2|} \quad as\ well\ as \quad \hat{H}_{ee} = \frac{e^2}{4\pi\varepsilon_0|r_{12}|} \tag{3}$$

The original formulation of the helium system has six degrees of freedom. Considering the spherical symmetry as described above we can reduce the formulation to $r_1$ and $r_2$, where $r_1 > 0, r_2 > 0$ are independent parameters. Consequently, the wave function we are looking for has therefore the form $R_{He}(r_1, r_2) = R_{He}(r_2, r_1)$.

In order to master the solution by facilitating a mathematical approach, we identify the function $R_{He}: \mathbb{R}^2 \to \mathbb{C}$ with its complex pendant $R_c: \mathbb{C} \to \mathbb{C}$, formally transforming $R_{He}(r_1, r_2)$ into $R_c(z)$ with $z = x + iy$ and interpreting $r_1$ as the real part of z and $r_2$ as its imaginary part. Thus, the substitution is like:

$$x = r_1 \quad and \quad y = -ir_2 \quad with \quad r_1, r_2 \in \mathbb{R}^+ \tag{4}$$

Note that $r_1, r_2$ stay real and positive to ensure $R_c(z) \to R_{He}(r_1, r_2)$ has plausible input parameters. The question arises how the radial wave function for two electrons can be rearranged in this way with $R_z: \mathbb{C} \to \mathbb{C}$ without the phase factor $e^{i\phi}$ in the first place? In the time-independent Schrödinger equation, the phase of the wave function describes the interference behavior of two states in superposition. For a single wave function, the phase multiplies out according to $|e^{i\phi}\psi|^2 = |\psi|^2$ and $\langle e^{i\phi}\psi|\hat{O}|e^{i\phi}\psi\rangle = \langle\psi|\hat{O}|\psi\rangle$, while $e^{i\phi}$ stays important in linear combinations causing interference effects. However, since the entangled electrons cannot sufficiently be described with a linear combination, we must cover them in a common wave function, making $e^{i\phi}$ irrelevant. Accordingly, the vacant parameter can be used for $r_2$.

The complex function $R_c(z)$ has two independent input parameters and is mathematically more elegant to handle than a function with two real parameters. This is particularly true as complex analysis offers various tools to calculate functions of this type. [12,13,55]

Since the momentum operator for both electrons must be well-defined over whole $\mathbb{C}$, $R_c$ must be holomorphic, i.e., complex differentiable, and this claim is even more powerful than calling for total differentiability of real functions. The technical details of the transformation and derivation of the expectation values using complex integration are lined out in the appendix (chapter 9).

## 4. Derivation of the general electron potential

Heisenberg's uncertainty principle is one of the most important foundations of quantum mechanics. Consequently, its formulation can be found in multiple relevant equations, be it in the equations of motion, or in the form of the commutation relation for observables. Consistently, however, this must also apply to the consideration of electrons, especially if these are implemented as contributors to potentials within the Schrödinger equation. From classical electrodynamics, the approach of the Coulomb potential for point charges has proven itself. Especially on the macroscopic charge and length scale, Coulomb's law is well documented in the literature [38,39,40]. However, if one considers relevant scales for quantum mechanics, e.g. individual electrons in an atom, then for potentials with an $\phi_C \propto r^{-1}$ characteristic, one would already be able to deduce the exact location of the electron by measuring its electric field gradient at few points, and according to Heisenberg this is not permitted without a diverging momentum. Consequently, the assumption frequently used in the literature that electrons in the atoms are modelled by point charges often leads to a dead end. Rather, we must realize that every point of the electron wave function manifests as a source of the electric potential, since the electron resides at that location with a certain probability, or under a different view i.e. integrated over time, spends a certain duration there, thereby generating a part of the potential. If we thus follow Heisenberg's requirement and utilize the approach of Green's function, the potential arises from the convolution of the Coulomb interaction with the charge density $\rho(\vec{r}')$, i.e., the square of the wave function:

$$\phi(\vec{r}) = \frac{1}{4\pi\varepsilon_0} \int \frac{\rho(\vec{r}')}{|\vec{r} - \vec{r}'|} d^3r' \tag{5}$$

The charge density $\rho(\vec{r}')$ now considers the spatial uncertainty, since the exact location of the electron and thus its field is smeared out. If we now look at the interaction term:

$$\hat{H}_{ee} = \frac{e^2}{4\pi\varepsilon_0 r_{12}} \quad with \quad r_{12} = |\vec{r} - \vec{r}'| = r^2 + r'^2 + 2rr'cos\theta \tag{6}$$

We see immediately that we can neither define the angle $\theta$ between the electrons, nor the positions $r'$ and $r$ precisely without violating Heisenberg's principle, so instead we use (5) to model the interaction.

At first glance, (5) resembles the Hartree mean field approach. However, in our approach, the wave function is not separated into an independent two-particle system in the form $R(r_1, r_2) = R_1(r_1)R_2(r_2)$, but both electrons are considered as entangled particles in a common wave function $R_c(z)$, and we calculate the self-interaction operator for $R_c(z)$, comparable to concepts used in the second quantization as in [82,83]. By the way, since Tolic has shown that an N-electron Schrödinger problem with pure Coulomb interactions is in fact mathematically unsolvable [81], it seems logical to engage the problem in an adapted way, namely with this approach.

In this context, we first consider the radial symmetry of the wave functions and switch from Cartesian coordinates to spherical coordinates. Since the $1s^2$singlet state is known to be spherical symmetric as discussed, we can restrict ourselves to the nuclear distances of the electrons $r_1, r_2$. To simplify further considerations, we realize the transition from $R_{He}(r_1, r_2)$ into $R_c(z)$ and, referring to the holomorphy of $R_c$ together with the calculation rules of chapter 9, we use $R_c(z)$ for integrals and differentials in the following calculations as needed. We understand (5) mathematically rigid as the convolution of the charge density $\rho$ with the potential $\phi_C$:

$$\phi(z) = [\rho * \phi_C](z) = \int_\gamma \rho(z')\phi_C(z - z')\, d^3z'$$

$$\Leftrightarrow \hat{H} = \int_\gamma \overline{R_c(z)} \left( \int_{\gamma'} \overline{R_c(z')} \frac{1}{z - z'} R_c(z') d^3z' \right) R_c(z) d^3z = \int_\gamma \overline{R_c(z)} \phi(z) R_c(z) d^3z \tag{7}$$

And the self-interaction term $\hat{H}$ (modulo constants) of $R_c$ with the kernel $\frac{1}{z-z'}$. Consequently, the expansion of the field origin from a singular point in space to an extended spatial zone has dramatic effects on the field. So, convolving the $r^{-1}$ field with an extended charge distribution providing exponential decay shows a decay of the field as well. As known, convolving a function with an integral kernel inherits the kernel properties to the function on its definition range.

Even more serious, however, is the fact that the field - and thus also the potential energy - no longer has a singularity, but remains continuous, differentiable and, above all, finite over the entire range, provided the charge density fulfills these criteria. In particular, the treatment of poles (e.g., through cusp conditions) can thereby be circumvented.

To correctly implement $\hat{H}_{ee}$ into the Schrödinger equation, the electron wave function must be made usable in generalized form as a potential term. This is particularly important because the potential terms must be evaluable in a differential equation and ultimately derive valid solutions. The general approach is therefore defined as a sum of arbitrary wave functions:

$$\rho(z) = z^2 R_c(z)^2 = \sum_n A_n z^n e^{-a_n|z|} \quad A_n \in \mathbb{R}, a_n > 0 \tag{8}$$

Remember, that $z^2$ is the volume element for the spherical coordinates. Given that $R_c(z)$ is holomorphic, square-integrable and normalized, and that $|R_c(z)|^2 > 0$ applies to whole $z \in \mathbb{R}_0^+ + i\mathbb{R}_0^+$, then analog to the solutions of the hydrogen atom, a general wave function can be specified as the sum of decreasing exponential functions multiplied by a polynomial as defined in (8). This applies, because $R_c(z)$ - due to holomorphy - can be always expanded locally into a Taylor power series converging against the function in a neighborhood of the development point. If we now use (8) considering a distributed charge, this also means that the convolution can be formed as an inverse Fourier transform from the multiplication of the two Fourier transforms according to the convolving integration theorem [24-28]. Note that the normalization factors are set to one due to convenient substitution. [27]

$$\phi(z) = [\rho * \phi_C](z) = \int_{\gamma'} \rho(z')\phi_C(z - z')\, dz' = \mathcal{F}^{-1}\left[\mathcal{F}(z^2|R_c(z)|^2) \cdot \mathcal{F}\big(\phi_C(z)\big)\right](z) \qquad (9)$$

And the Fourier transform functions respectively:

$$\mathcal{F}\big(\phi_C(z)\big)(\omega) = \mathcal{F}\left(\frac{q}{4\pi\varepsilon_0}\frac{1}{|z|}\right)(\omega) = -i\frac{q}{4\pi\varepsilon_0}\, sgn(\omega) \qquad (10)$$

$$\mathcal{F}\big(\rho(z)\big)(\omega) = \mathcal{F}(z^2|R_c(z)|^2)(\omega) \qquad (11)$$

Note that even if we focus on $\mathbb{R}_0^+ + i\mathbb{R}_0^+$ considering the wave function, the Fourier transformation must be valid on $\mathbb{C}$. If we now apply the general formulation for a charge distribution from (8), the expression can be converted into a Fourier transform:

$$\mathcal{F}\left(\sum_n A_n z^n e^{-a_n|z|}\right)(\omega) = \sum_n \frac{A_n n!}{(a_n + i\omega)^{n+1}} \quad with \;\; A_n, a_n \in \mathbb{R}, a_n > 0 \qquad (12)$$

(12) is only plausibly defined $|z| > 0$. Combined with the Fourier transform of the potential, it results in:

$$\phi(z) = \frac{q}{4\pi\varepsilon_0}\mathcal{F}^{-1}\left\{\mathcal{F}\left(\frac{1}{|z|}\right)\mathcal{F}\left(\sum_n A_n z^n e^{-a_n|z|}\right)\right\}(z) \qquad (13)$$

This leads to the fully formulated integral:

$$\Leftrightarrow \phi(z) = -i\frac{q}{4\pi\varepsilon_0}\sum_n \int_{-\infty}^{\infty} \frac{A_n n!\, sgn(\omega) e^{i\omega z}}{(a_n + i\omega)^{n+1}} d\omega \qquad (14)$$

$$\Leftrightarrow \phi(z) = -i\frac{q}{4\pi\varepsilon_0}\sum_n A_n n! \lim_{b\to 0}\left[-\int_{-\infty}^{\infty} \frac{u(-\omega)e^{b\omega}e^{i\omega z}d\omega}{(a_n + i\omega)^{n+1}} + \int_{-\infty}^{\infty} \frac{u(\omega)e^{-b\omega}e^{i\omega z}d\omega}{(a_n + i\omega)^{n+1}}\right] \qquad (15)$$

The definition of the sign-function was utilized here. Also note that the imaginary value can be mapped to the real axis by phase rotation.

$$sgn(\omega) = \lim_{b\to 0} u(-\omega)e^{b\omega} + \lim_{b\to 0} u(\omega)\, e^{-b\omega} \; and \; u(\omega) = \begin{cases} 1 & \omega > 0 \\ 0 & \omega < 0 \end{cases} \qquad (16)$$

By integration the following expression is obtained.

$$\Leftrightarrow \phi(z) = \frac{q}{4\pi\varepsilon_0}\sum_n \Bigg(-\frac{A_n}{2}\lim_{b\to 0}\left[\Theta(|z| + b)(|z| + b)^n e^{-a_n(-|z|-b)}\right]_{-\infty}^{\infty} + \frac{A_n}{2}\lim_{b\to 0}\left[\Theta(|z| - b)(|z| - b)^n e^{-a_n(|z|-b)}\right]_{-\infty}^{\infty}\Bigg) \qquad (17)$$

Now swap the integration limits on the left and substitute with $-|z|$

$$\Leftrightarrow \phi(z) = \frac{q}{4\pi\varepsilon_0}\sum_n \left(\frac{A_n}{2}\lim_{b\to 0}\left[\Theta(|z|+b)(|z|+b)^n e^{-a_n(-|z|-b)}\right]_{-\infty}^{\infty} + \frac{A_n}{2}\lim_{b\to 0}\left[\Theta(|z|-b)(|z|-b)^n e^{-a_n(|z|-b)}\right]_{-\infty}^{\infty}\right) \tag{18}$$

$$\Leftrightarrow \phi(z) = \frac{q}{4\pi\varepsilon_0}\sum_n \int_{-\infty}^{\infty} \frac{A_n n!}{(a_n + i\omega)^{n+1}} sgn(\omega) e^{i\omega z} d\omega = \frac{q}{4\pi\varepsilon_0}\sum_n A_n z^n e^{-a_n|z|} \tag{19}$$

The wave function $R_c(z)$ covers an entangled state for $Z$=2 electrons, therefore $q = Ze$. It is also important to note that the Schrödinger equation does not refer to the potential $\phi$, but the potential energy $V$, i.e., this is obtained by multiplying $\phi$ by $e$:

$$V(z) = e\phi(z) = \frac{Ze^2}{4\pi\varepsilon_0}\sum_n A_n z^n e^{-a_n|z|} \tag{20}$$

Thus (20) results in a formulation of a radial symmetric electron potential energy, which can be processed and implemented in the Schrödinger equation. $V(z)$ describes the situation, that every potential caused by electrons present in the vicinity of the atom can be modelled as a sum over $n$ of a polynomial expression multiplied with an exponential damped probability depending on the distance to the core with the specific parameters $A_n$ and $a_n$.

To model the electron potential correctly, we must consider, that the electrons interact with their electrical potential by itself and mutually, i.e. for electron $i, j$ we consider four interaction terms: $i \leftrightarrow i$ , $j \leftrightarrow j$ as well as $i \leftrightarrow j$ and $j \leftrightarrow i$. So, the specific formula of $V_E(z)$ used here is therefore:

$$V_E(z) = 4\frac{Ze^2}{4\pi\varepsilon_0}\sum_n A_n z^n e^{-a_n|z|} = 4\frac{Ze^2}{4\pi\varepsilon_0}\rho(z) = 4\frac{Ze^2}{4\pi\varepsilon_0} z^2 R_c(z)^2 \tag{21}$$

The expression $V_E(z)$ can now be implemented in the Schrodinger equation.

## 5. The analytical solution of the Schrödinger equation for Helium

Finally, we need to solve the Schrödinger equation. To start according to the literature, the original Hamiltonian $\widehat{H}$ of the helium system would look like:

$$\widehat{H}|\psi,\phi\rangle = E|\psi,\phi\rangle \quad with \quad \widehat{H} = \widehat{H}_1 + \widehat{H}_2 + \widehat{H}_{ee} \quad and \tag{22}$$

$$\widehat{H}_1 = \frac{\hat{p}_1{}^2}{2m_e} - \frac{Ze^2}{4\pi\varepsilon_0 r_1} \quad and \quad \widehat{H}_2 = \frac{\hat{p}_2{}^2}{2m_e} - \frac{Ze^2}{4\pi\varepsilon_0 r_2} \quad as\ well\ as \quad \widehat{H}_{ee} = \frac{Ze^2}{4\pi\varepsilon_0 r_{12}} \tag{23}$$

as mentioned in chapter 3 with $Z$ defined as the charge number. In contrast to the literature, a significantly different approach is necessary here to advance to an analytic solution. We motivated already the use of $R_c(z)$ as the wave function for both entangled electrons and laid the foundation to do the calculations. As $R_c(z)$ refers to both electrons, $\widehat{H}_1 + \widehat{H}_2$ must be regarded as one operator. Furthermore, $\widehat{H}_{ee}$ is transformed to $\widehat{H}_{ee}{}^z$ according to (21)

$$\widehat{H}_{ee} = \frac{e^2}{2\pi\varepsilon_0}\frac{1}{r_{12}} \curvearrowright \widehat{H}_{ee}{}^z = 4\frac{Ze^2}{4\pi\varepsilon_0}z^2 R_c(z)^2 \tag{24}$$

This put together results in a Schrödinger equation for two electrons. We use the absolute value of the total energy $|E|$ to evaluate the collective ground state energy of the electrons in (25).

As we have shown in chapter 3 the helium ground state is spherically symmetric, so by utilizing especially $R_c(z)$ with its holomorphic characteristics, we stay with a spherical symmetric coordinate system, but we are using a complex wave function. So, this results in the Hamiltonian $\widehat{H}_c$:

$$\widehat{H}_c = \frac{\hat{p}(z)^2}{2m_e} - \frac{Ze^2}{4\pi\varepsilon_0 z} + \frac{2\hbar^2 L(L+1)}{2m_e z^2} + 4\frac{Ze^2}{4\pi\varepsilon_0}z^2 R_c(z)^2 \tag{25}$$

Fully written out this results in the equation:

$$\frac{2}{z^2}\frac{\partial}{\partial z}\left(z^2\frac{\partial}{\partial z}R_c(z)\right) + \frac{2m_e}{\hbar^2}\left(\frac{2Ze^2}{4\pi\varepsilon_0}\frac{1}{z} - \frac{2\hbar^2 L(L+1)}{2m_e z^2} - 4\frac{Ze^2}{4\pi\varepsilon_0}z^2 R_c{}^2(z) - |E|\right)R_c(z) = 0 \tag{26}$$

$$\Leftrightarrow \frac{\partial^2 R_c(z)}{\partial z^2} + \frac{2}{z}\frac{\partial R_c(z)}{\partial z} + \frac{2m_e}{\hbar^2}\left(\frac{Ze^2}{4\pi\varepsilon_0}\frac{1}{z} - \frac{\hbar^2 L(L+1)}{2m_e z^2} - 4\frac{Ze^2}{4\pi\varepsilon_0}z^2 R_c{}^2(z) - |E|\right)R_c(z) = 0 \tag{27}$$

Solving (27) generally is a particularly challenging task due to the non-linear potential term of the electrons. We focus only on the states for the two-electron atom where $L = 0$ is considered, so we simplify (27) while also utilizing the following substitutions simultaneously:

$$A \equiv \frac{2m_e}{\hbar^2}\frac{Ze^2}{4\pi\varepsilon_0}, \qquad B \equiv \frac{2m_e}{\hbar^2}, \qquad C_e \equiv \frac{2m_e}{\hbar^2}4\frac{Ze^2}{4\pi\varepsilon_0} \tag{28}$$

Accordingly, (27) results in a differential equation for the $l = 0$ states of helium as a two-electron system:

$$\Leftrightarrow \frac{\partial^2 R_c(z)}{\partial z^2} + \frac{2}{z}\frac{\partial R_c(z)}{\partial z} + \left(\frac{A}{z} - B|E|\right)R_c(z) - C_e z^2 R_c(z)^2 R_c(z) = 0 \tag{29}$$

For further work, it has proven to be useful to transform the differential equation into Laplace space [24,51,52,53,56,57,58]. The Laplace transformation itself is complex, and it requires integrable functions with lower growth than $e^{-st}$ for $t \longrightarrow +\infty$. It can also be expanded to a complex input variable $t \longrightarrow z$ if we use a proper path $\gamma$ for integration and a well-defined function. As $R_c(z)$ $is$ holomorphic and the properties of chapter 9 are valid, we may proceed. However, reciprocal z-terms cause difficulties during the transformation, which is why (29) is multiplied by $z \neq 0$ beforehand:

$$\Leftrightarrow z\frac{\partial^2 R_c(z)}{\partial z^2} + 2\frac{\partial R_c(z)}{\partial z} + (A - zB|E|)R_c(z) - z\,C_e z^2 R_c(z)^2 R_c(z) = 0 \tag{30}$$

Now translate (30) into the Laplace transform according to the conversion rules from [56,57]. Let $\mathfrak{R}_c(s) = \mathcal{L}_z[R_c(z)](s)$ be the Laplace transform of the wave function. In addition,

$$\mathcal{L}_z[z^2 R_c(z)^2](s) \equiv \mathfrak{f}_z(s) \tag{31}$$

should be defined to facilitate processing later. This results in the Laplace-transformed Schrödinger equation as:

$$\begin{aligned} -s^2\frac{\partial}{\partial s}\mathfrak{R}_c(s) - 2s\mathfrak{R}_c(s) + R_c(0) + 2\big(s\mathfrak{R}_c(s) - R_c(0)\big) + A\mathfrak{R}_c(s) + B|E|\frac{\partial}{\partial s}\mathfrak{R}_c(s) \\ + C_e\mathfrak{f}_z * \frac{\partial}{\partial s}\mathfrak{R}_c(s) = 0 \end{aligned} \tag{32}$$

Note, that $C_e\mathfrak{f}_z * \frac{\partial}{\partial s}\mathfrak{R}_c(s)$ is the convolution of both functions in Laplace space. First one ought to simplify the equations:

$$\Leftrightarrow -(s^2 - B|E|)\frac{\partial}{\partial s}\mathfrak{R}_c(s) + A\mathfrak{R}_c(s) - R_c(0) + C_e\mathfrak{f}_z * \frac{\partial}{\partial s}\mathfrak{R}_c(s) = 0 \tag{33}$$

To continue, theorems from functional analysis are referred. One important is the theorem of Fubini and Tonelli which states that the finite integral of a convolution of two non-negative functions is equal to the multiplication of the two function integrals [43-45,48,49,50,54]. For our case, this means:

$$\int_\gamma [\mathfrak{F}(s) * \mathfrak{G}(s)]ds = \int_\gamma \mathfrak{F}(s)ds \int_\gamma \mathfrak{G}(s)ds \tag{34}$$

So, using this on the convolution in 33, we can simplify $\mathfrak{f}_z$ into a handier term by integration:

$$\int_\gamma \mathfrak{f}_z ds = \int_\gamma \mathcal{L}_z[z^2 R_c(z)^2](s) ds = \int_\gamma \left( \int_\gamma z^2 R_c(z)^2 \, e^{-sz} dz \right) ds$$
$$= \int_\gamma \left( \int_\gamma z^2 R_c(z)^2 \, e^{-sz} ds \right) dz = \int_\gamma z^2 R_c(z)^2 \left( \int_\gamma e^{-sz} ds \right) dz \tag{35}$$
$$= \frac{1}{s} \int_\gamma z^2 R_c(z)^2 dz = \frac{1}{s} \quad with \quad \int_\gamma z^2 R_c(z)^2 dz \equiv 1$$

We can now utilize (35) on $C_e \mathfrak{f}_z * \frac{\partial}{\partial s} \mathfrak{R}_c(s)$ and applying the identity $\frac{\partial}{\partial s} \int ds = \mathbb{1}$ like this:

$$\mathfrak{f}_z * \frac{\partial}{\partial s} \mathfrak{R}_c = \frac{\partial}{\partial s} \int_\gamma \left[ \mathfrak{f}_z * \frac{\partial}{\partial s} \mathfrak{R}_c \right] ds = \frac{\partial}{\partial s} \int_\gamma \mathfrak{f}_z ds \int_\gamma \frac{\partial}{\partial s} \mathfrak{R}_c ds = -\frac{1}{s^2} \mathfrak{R}_c \tag{36}$$

We apply (36) to (33) and get the simplified result:

$$\Leftrightarrow -(s^2 - B|E|) \frac{\partial}{\partial s} \mathfrak{R}_c(s) - R_c(0) + A \mathfrak{R}_c(s) - C_e \frac{1}{s^2} \mathfrak{R}_c = 0 \tag{37}$$

We can now rearrange the equation:

$$\Leftrightarrow \frac{\partial}{\partial s} \mathfrak{R}_c(s) = + \frac{A}{s^2 - B|E|} \mathfrak{R}_c(s) - \frac{1}{s^2 - B|E|} R_c(0) - \frac{C_e}{s^2 (s^2 - B|E|)} \mathfrak{R}_c \tag{38}$$

For unification reasons let us define:

$$\alpha_0{}^2 = B|E| \quad with \quad \alpha_0 \in \mathbb{R}^+ \tag{39}$$

Since we use the absolute value of binding energy the root remains real. The inverse transform to the spatial $z$-dimension is once again straight forward:

$$-z R_c(z) = + \frac{A}{\alpha_0} \big( sinh(\alpha_0 z) * R_c(z) \big) - \frac{R_c(0)}{\alpha_0} sinh(\alpha_0 z)$$
$$- \frac{C_e}{\alpha_0{}^3} \Big( \big( \alpha_0 z - sinh(\alpha_0 z) \big) * R_c(z) \Big) \tag{40}$$

Remember, that by back transformation into z-space the product of two functions in Laplace space $\mathfrak{F}(s)\mathfrak{G}(s)$ changes into the convolution $F(z) * G(z)$ in z-space. So, first converting the left side of the equation in a straight term reveals:

$$R_c(z) = -A \frac{sinh(\alpha_0 z)}{\alpha_0 z} * R_c(z) + R_c(0) \frac{sinh(\alpha_0 z)}{\alpha_0 z} + \frac{C_e}{\alpha_0{}^2} \left( 1 - \frac{sinh(\alpha_0 z)}{\alpha_0 z} \right) * R_c(z) \tag{41}$$

Again, the theorem of Fubini and Tonelli can be used analogue to (36), and the result is obtained by applying the identity $\frac{\partial}{\partial z} \int dz = \mathbb{1}$ in complex $z$-plane. We use two steps, first writing down the integrals:

$$\int_\gamma R_c(z)\,dz = -A\int_\gamma \frac{sinh(\alpha_0 z)}{\alpha_0 z}dz\int_\gamma R_c(z)\,dz + R_c(0)\int_\gamma \frac{sinh(\alpha_0 z)}{\alpha_0 z}dz + \frac{C_e}{{\alpha_0}^2}\int_\gamma\left(1-\frac{sinh(\alpha_0 z)}{\alpha_0 z}\right)dz\int_\gamma R_c(z)\,dz \tag{42}$$

And then we can now rearrange the equation:

$$\int_\gamma R_c(z)\,dz\left(1 + A\int_\gamma \frac{sinh(\alpha_0 z)}{\alpha_0 z}dz - \frac{C_e}{{\alpha_0}^2}\int_\gamma\left(1-\frac{sinh(\alpha_0 z)}{\alpha_0 z}\right)dz\right) = R_c(0)\int_\gamma \frac{sinh(\alpha_0 z)}{\alpha_0 z}dz \tag{43}$$

Resulting in the equation (44):

$$\Leftrightarrow \int_\gamma R_c(z)\,dz = R_c(0)\frac{\int \frac{sinh(\alpha_0 z)}{\alpha_0 z}dz}{1 + A\int \frac{sinh(\alpha_0 z)}{\alpha_0 z}dz - \frac{C_e}{{\alpha_0}^2}\int\left(1-\frac{sinh(\alpha_0 z)}{\alpha_0 z}\right)dz} \tag{44}$$

And now we generate $R_c(z)$ by building the derivative and applying $\frac{\partial}{\partial z}$ on both sides of the equation:

$$R_c(z) = R_c(0)\frac{\left(1-\frac{C_e}{{\alpha_0}^2}z\right)\frac{sinh(\alpha_0 z)}{\alpha_0 z} + \frac{C_e}{{\alpha_0}^2}\int \frac{sinh(\alpha_0 z)}{\alpha_0 z}dz}{\left(1 + A\int \frac{sinh(\alpha_0 z)}{\alpha_0 z}dz - \frac{C_e}{{\alpha_0}^2}\int\left(1-\frac{sinh(\alpha_0 z)}{\alpha_0 z}\right)dz\right)^2} \tag{45}$$

This is a rather unhandy result for further use. But analyzing (45) closer and taking (52) into account, we find that $\frac{C_e}{{\alpha_0}^2} \approx 10^{-10} \ll 1$ and thus, the concatenated terms can be neglected without falsifying the result. The approximation results in equation (46):

$$\frac{C_e}{{\alpha_0}^2} \ll 1 \quad \Leftrightarrow \quad R_c(z) = R_c(0)\frac{sinh(\alpha_0 z)}{\alpha_0 z\left(1 + A\int \frac{sinh(\alpha_0 z)}{\alpha_0 z}dz\right)^2} \tag{46}$$

Or reformulated with (48), where $Shi(\alpha_0 z)$ is the hyperbolic sine integral function which has no elementary form, we get:

$$\Leftrightarrow R_c(z) = R_c(0)\frac{sinh(\alpha_0 z)}{\alpha_0 z\big(1 + AShi(\alpha_0 z)\big)^2} \tag{47}$$

$Shi(\alpha_0 z)$ is formally defined as:

$$Shi(\alpha z) = \int_0^z \frac{sinh(\alpha\xi)}{\alpha\xi}d\xi \tag{48}$$

and can be written by the Taylor series expansion:

$$Shi(z) = z + \frac{z^3}{3*3!} + \frac{z^5}{5*5!} + \frac{z^7}{7*7!} + \ldots \quad (49)$$

which converges for all complex values $z$ and describes an entire transcendental function. $R_c(z)$ is the analytically derived radial wave function for helium and helium-like systems in the ground state for two electrons with *n=1, S=0, L=0*. We normalize the wave function by $f_N$. This was neglected until now for convenience but is of course important for accurate calculations. We obtain:

$$R_c(z) = f_N \frac{sinh(\alpha_0 z)}{\alpha_0 z\big(1 + AShi(\alpha_0 z)\big)^2} = f_N \frac{sinh\big(\alpha_0(x+iy)\big)}{\alpha_0(x+iy)\Big(1 + AShi\big(\alpha_0(x+iy)\big)\Big)^2} \quad (50)$$

With $z = x + iy$ and

$$\int_0^\infty z^2 R_c(z)^2 dz \equiv \frac{1}{{f_N}^2}, \qquad f_N > 0 \quad (51)$$

(50) and (51) are covered by the formula of the general electron potential proven in chapter 4 and 9. It should also be mentioned that there was not any analytical root function found for the integral to determine $f_N$, thus the integral is integrated numerically. Resubstitute with (39) then one finds:

$${\alpha_0}^2 = B|E| = \frac{2m_e}{\hbar^2}|E_0| \quad (52)$$

We can interpret $\alpha_0$ as the decay rate of the radial helium wavefunction. The calculations shown in chapter 7 deliver:

$$\boldsymbol{\alpha_0 = 4.55380987904454681(72) * 10^{10}\, m^{-1}}$$

If we look at the behavior at $z \to 0$ and $z \to \infty$, we see that $R_c$ converges to $\lim_{z\to 0} R_c(z) \to f_N$. For $\lim_{z\to\infty} R_c(z)$ we find the limit behavior with the identity $sinh(\alpha_0 z) = \frac{1}{2}(e^{\alpha_0 z} - e^{-\alpha_0 z})$:

$$\begin{aligned} Shi(\alpha_0 z) &= \int_0^z \frac{sinh(\alpha_0 \xi)}{\alpha_0 \xi} d\xi = \int_0^z \frac{e^{\alpha_0 \xi} - e^{-\alpha_0 \xi}}{2\alpha_0 \xi} d\xi \\ &\xrightarrow{z\to\infty} \frac{1}{2} Ei(\alpha_0 z) \to e^{\alpha_0 z}\left[\frac{1}{2\alpha_0 z} + \mathcal{O}\left(\frac{1}{(\alpha_0 z)^2}\right)\right] \end{aligned} \quad (53)$$

This results in a behavior:

$$\lim_{z\to\infty} R_c(z) = f_N \frac{e^{\alpha_0 z}}{\alpha_0 z\left(1 + Ae^{\alpha_0 z}\left(\frac{1}{2\alpha_0 z} + \mathcal{O}\left(\frac{1}{(\alpha_0 z)^2}\right)\right)\right)^2} \to f_N \frac{e^{\alpha_0 z}}{\frac{A^2}{4\alpha_0}\frac{1}{z}e^{2\alpha_0 z}} \to f_N \frac{4\alpha_0}{A^2} z e^{-\alpha_0 z} \quad (54)$$

So, for large z we find a behavior of $\lim_{z\to\infty} R_c(z) \propto e^{-\alpha_0 z}$.

## a. *Interpretation of the two-electron concept of $R^n{}_{He}$ and comparison with the Hylleraas approach*

The work on solving the Schrödinger equation for the helium atom now spans almost a century since the beginnings of quantum mechanics and is so extensive and diverse that it is impossible to cover it exhaustively in this short article. We can therefore only review a few limited prominent examples and refer primarily to the original work of Hylleraas, which has significantly influenced further research.

Using complex analysis shows being convenient, but the question arises how the holomorphic wave function $R_c$ is to be understood as a function of two electrons.

$$R_c(z) = f_N \frac{sinh\big(n\alpha_0(x+iy)\big)}{\alpha_0(x+iy)\Big(1+AShi\big(n\alpha_0(x+iy)\big)\Big)^2} \quad with \quad Re\{z\}, Im\{z\} > 0 \tag{55}$$

The complex variable z can be transferred back into two independent parameters $r_1$ and $r_2$. The wave function $R_c \longrightarrow R_{He}$ can be written by using (4): $x = r_1, y = -ir_2$

$$R_{He}(r_1, r_2) = f_N \frac{sinh\big(n\alpha_0(r_1+r_2)\big)}{\alpha_0(r_1+r_2)\Big(1+AShi\big(n\alpha_0(r_1+r_2)\big)\Big)^2}, \qquad r_1, r_2 \geq 0 \tag{56}$$

Considering the wave function, provided by Hylleraas, it is written as:

$$\Psi(s,t,u) = exp(-ks)\sum_{lmn}^{N} c_{l,2m,n} s^l t^{2m} u^n \tag{57}$$

With the following properties and $r_{12}$ as defined:

$$s = r_1 + r_2, \qquad t = r_1 - r_2, \qquad u = r_{12} \tag{58}$$

Remember, that we are looking de facto at an infinite series, thus $s^l t^{2m} u^n$ being the elementary terms with the coefficients $c_{l,2m,n}$ respectively. The coefficients are then optimized by variation calculation to find the energy minimum. Already a few parameters give a very high precision [4].

The terms $exp(-ks)$ and $s^l$ seem familiar if one looks at the Taylor expansion of $sinh(\alpha_0 z)$. The term $t^{2m}$ is difficult to be assigned and $u^n$ is missing completely in (56), especially due to the topics discussed in chapter 4. One reason for a high number of summators to reach an adequate precision for the energy may lie in the fact, that the wave function $R_c$ or $R_{He}$ is an entire transcendental function, thus not trivially reproducible by elementary functions like $r^n exp(-kr)$, but very well replicable by a series with a sufficient number of terms. Especially the Shi-function tuned with the factor $A$ is deviating significantly from elementary functions making the modelling by series expansions extensive.

Despite the enormous success of Hylleraas' and many of the other numerical methods, its greatest disadvantage from the author's point of view lies in the fact that only a limited gain in knowledge is possible through the series development. Nevertheless, the use of series connected with the calculus of variations is a particularly important building block for the modeling of quantum mechanical systems.

### b. *Comparison with the Schrödinger solution of the Hydrogen atom*

A direct comparison with the original hydrogen solution of the Schrödinger equation reveals interesting differences. To visualize this, both wavefunctions and the radial probability density are shown in figure 1 and 3, respectively. Figure 2 shows $R_{He}$ in the two-dimensional plane for visualization. So, if we look at the radial probability density for both cases, one notifies:

$$\left|R_{1,0,0}^{H}(r)\right|^2 = \frac{4Z^3}{a_0{}^3} e^{-\frac{4r}{a_0}} \tag{59}$$

And with (56) we obtain:

$$R_{He}(r_1, r_2)^2 = f_N{}^2 \frac{sinh^2\big(\alpha_0(r_1 + r_2)\big)}{\alpha_0{}^2|r_1 + r_2|^2 \Big(1 + AShi\big(\alpha_0(r_1 + r_2)\big)\Big)^4} \quad r_1, r_2 \geq 0 \tag{60}$$

With the Bohr radius $a_0 = \frac{4\pi\varepsilon_0\hbar^2 n^2}{Ze^2 m_e}$ and the decay rate $\alpha_0$ for the helium solution, we can compute the maximum of the electron probability density

$$\boldsymbol{a_0{}^{He} = 1.58219417134664422(78) * 10^{-11}\, m}$$

for the electrons, which is consistent to literature values of the covalent radius for helium [59-70]. For $r_1, r_2$ one would find the maximum at $r_{max} = (t-1)r_1 + tr_2, t \in [0,1]$. That is a factor 3.345 smaller than the hydrogen solution. Together with the ratio of the spatial decay rates

$$\boldsymbol{\alpha_0 a_0 = 2.409772410789371 60(43)}$$

which is higher for $R^1{}_{He}$, thus, the neutral helium is much more compact than the hydrogen atom. Obviously the two electrons are located very closely to the core in He with a maximum at ≈15.8 pm and cause a higher electrical shielding thus making helium a tiny atomic system that appears neutral already in distances up to the high picometer range.

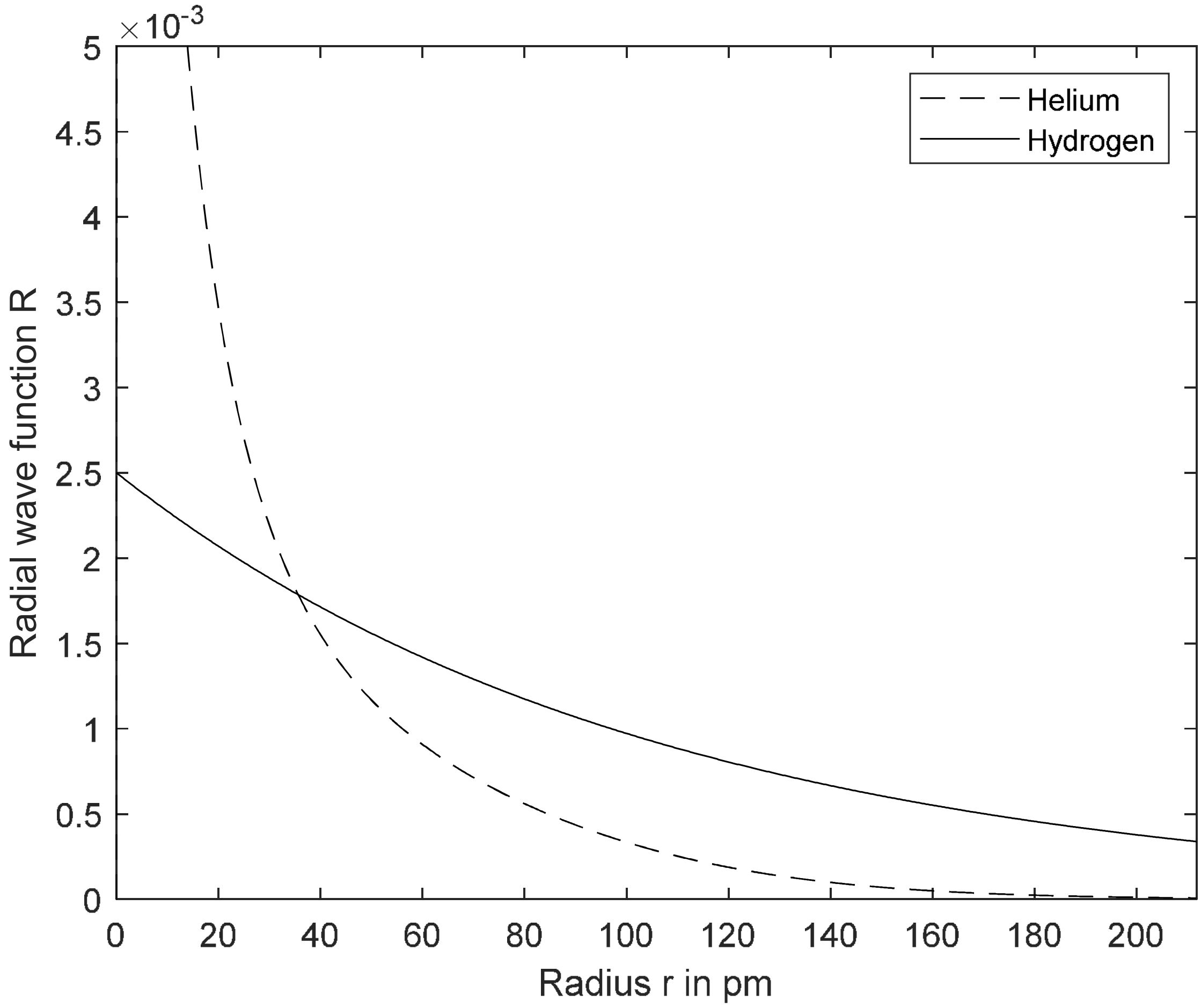

Fig.1 Wave functions of the helium atom (dashed line) and the hydrogen atom (full line). The helium wave function is plotted in one variable $r_i$ for one of the electrons with the other variable $r_{1-i}$ set to 0. Note that $R_{He}$ looks similar to the hydrogen solution of the Schrödinger equation, only the profile differs to a simple exponential decay manifesting an entire transcendental function. $R_{He}$ starts on a much higher level at the origin, though is finite at $r_i = 0$ and has a significantly steeper decay over $r_i$.

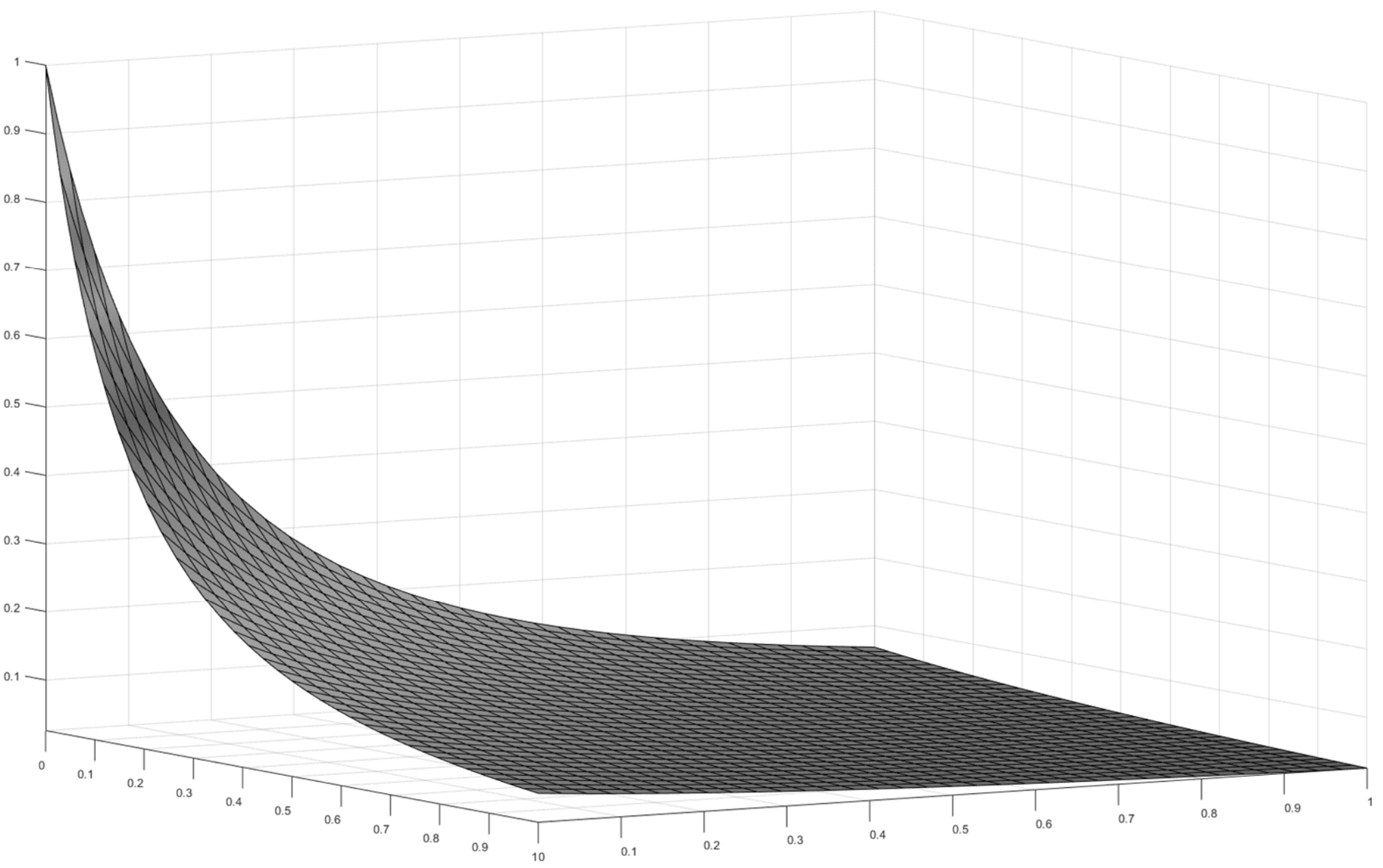

Fig.2 Function plot of $\frac{sinh(r_1+r_2)}{(r_1+r_2)\left(1+Shi(r_1+r_2)\right)^2}$ with arbitrary values as drawn in absolute values in z-direction over the $r_1 r_2$-plane and $A = \alpha_0$ set to 1. One can get an impression of the symmetric characteristics of $R_{He}(r_1, r_2)$. It can be identified easily that a transformation from $r_1 \rightarrow r_2$ and vice versa can be accomplished by a clockwise or counter-clockwise rotation by 90° in the plane.

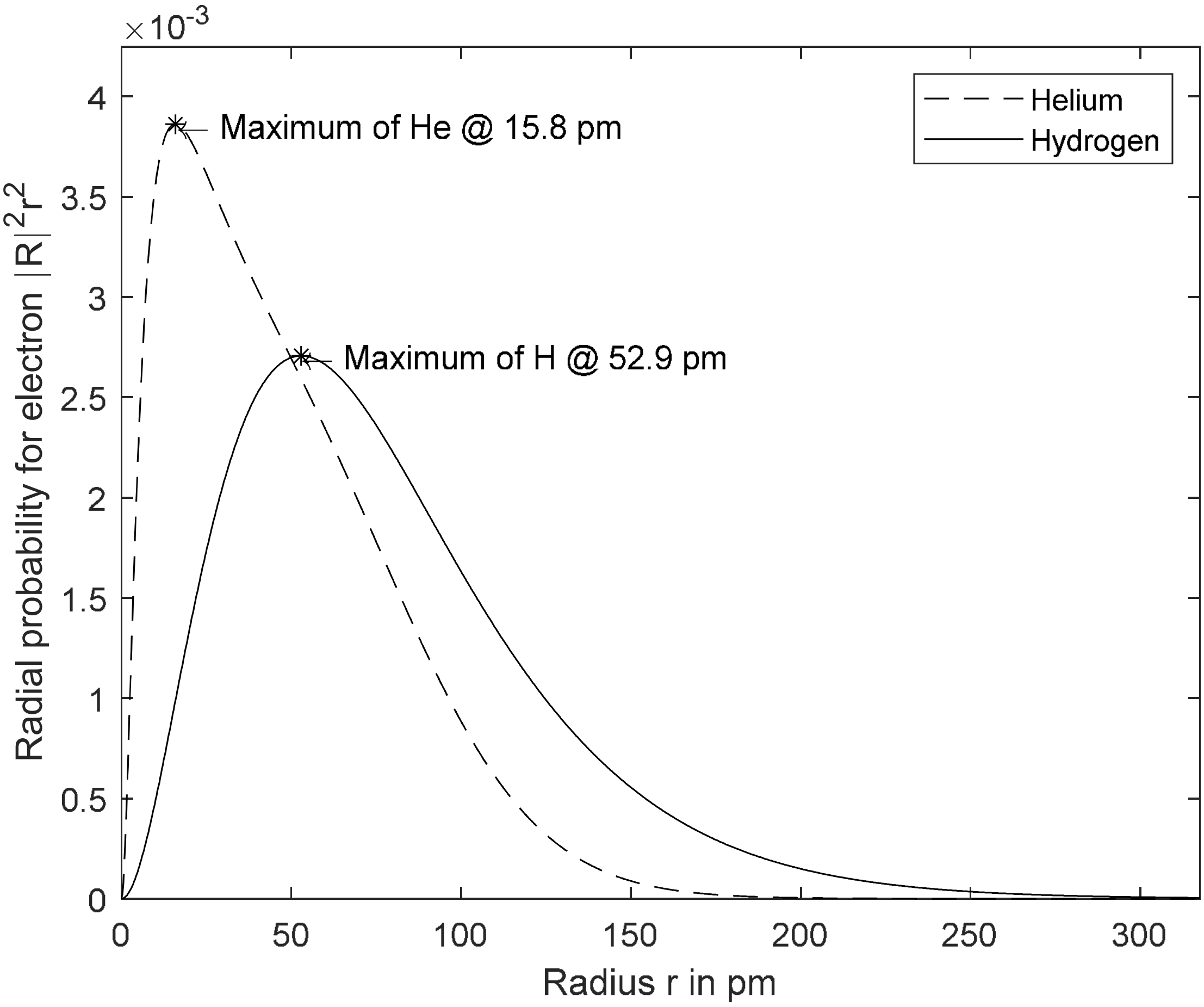

Fig.3 Radial probability density for the helium ground state (dashed line) and the hydrogen ground state (full line). The probability density for helium is again plotted for one variable $r_i$ with the other set to 0. Note that the maximum of the H-probability function is at 52.9 pm, as known from the literature, while the electrons in helium have their maximum probability density at 15.8 pm. Consequently, helium shows a much higher charge density closer to the core. Hence the neutral helium atom appears much smaller than hydrogen.

## 6. Corrective terms for the precise evaluation of the ground state energy

To keep an overview of the numerous correction terms, we introduce and motivate them shortly in this chapter and punctually show results if helpful in the context. A complete overview of the results will be given in the next chapter based on the calculations of the MATLAB script in chapter 12. Though the fundamental Hamiltonian with its resulting wave function in chapter 5 already produces decent results with $|E_0| \approx 80.0101\ eV$, it is possible to improve the accuracy by considering perturbations like the Breit operator or the Lamb shift. Additional to well-known perturbations we propose an energy contribution by a vibrational ground mode. This mechanism is of course familiar in molecular systems but rather unusual and surprising in neutral atoms: In this case it gives the electrons the ability to generate a dark vibrational mode.

### *a. Energy contributions by vibrational electronic modes*

If we analyze the potential terms of the Hamiltonian in (30) we find the expression:

$$\widehat{H}_c{}^{pot} = \frac{Ze^2}{4\pi\varepsilon_0}\left(4z^2 R_c(z)^2 - \frac{1}{z}\right)$$

Looking at the expectation value of the energy, we find a quasi-harmonic potential, that indicates the presence of vibrational modes. In fact, in contrast to Hydrogen, where the electron is coupled to the electromagnetic field due to spin - and angular momentum in case of excitations, the helium system does not have such a coupling in the ground state, as *S*=*L*=0. So, we can expect that collective vibrational modes of the electron do not interact with the electromagnetic field directly i.e. generating a dark vibrational mode, that can be excited by thermal or scattering interactions. If we consider both electrons as one system with two vibrational modes, swinging parallel or antiparallel to the radius, then the energy of this harmonic oscillator equals [36,37,41,42,47,]:

$$\langle E_{vib}\rangle = \hbar\omega\left(n + \frac{1}{2}\right) \quad with \quad \langle E_{vib}\rangle_0 = \frac{1}{2}\hbar\sqrt{\frac{k}{m_e}} \tag{61}$$

With $\langle E_{vib}\rangle_0$ being the zero point energy for each electron. The spring parameter $k$ can be derived by the second derivative of the potential energy at the point of maximum probability density of the electron $a_0{}^{He}$, i.e.:

$$k \equiv \frac{\partial^2}{\partial z^2}\langle E_{pot}\rangle\bigg|_{a_0{}^{He}} = \frac{\partial}{\partial z}\left(z^2\frac{A}{z}R_c{}^2 - C_E z^4 R_c{}^4\right)\bigg|_{a_0{}^{He}} \tag{62}$$

It can be easily seen that an antiparallel mode $\frac{\partial^2}{\partial x^2}\langle E_{pot}\rangle - \frac{\partial^2}{\partial y^2}\langle E_{pot}\rangle$ will vanish, so we concentrate on the parallel mode. The MATLAB file in chapter 12 uses the calculation in (62) to derive the perturbation energy. We get a ground energy of ≈ 0.6982 eV and the first excitation energy of ≈ 2.0946 eV, that would have an equivalent activation temperature of $\frac{1}{k_B}\langle E_{vib}\rangle_{n=1} \approx$ 24,307 K.

This approach could also give insight in the chemical properties of helium: Energy can be absorbed by vibrational modes without changing the electronic configuration, until sufficient activation energy for excitation or the vacuum state is provided. Smaller amounts of energy as disturbances from scattering or heat result in a transient buffering without the atom taking part in any further chemical reaction. Therefore, this mechanism could contribute to the stability of helium and the fundamental inertia mechanism of closed-shell systems, as typical for noble gases.

### *b. Relativistic kinetic energy correction*

The relativistic correction to the kinetic energy arises from the Taylor expansion of the energy-momentum relation [40-42]:

$$E = mc^2 + \frac{p^2}{2m_e} - \frac{p^4}{8{m_e}^3c^2} + \cdots$$

The third term provides the first relativistic correction to the non-relativistic kinetic energy. The corresponding Hamiltonian would then be:

$$H_R = -\frac{\hbar^4}{8{m_e}^3c^2}(\nabla^2)^2 \tag{63}$$

Where $(\nabla^2)^2$ is the Bi-Laplace operator in spherical coordinates acting on the wave function. For a spherically symmetric function $\Phi(r) = r^2\,\mathrm{f}(r)^2$, this results in

$$(\nabla^2)^2\Phi(r) = 6f(r)^2 + 4rf(r)\frac{\partial}{\partial r}f(r) + 2r^2\left(\frac{\partial}{\partial r}f(r)^2 + f(r)\frac{\partial^2}{\partial r^2}f(r)\right) \tag{64}$$

### *c. The Darwin term*

The Darwin term emerges from the Foldy–Wouthuysen transformation of the Dirac Hamiltonian. It accounts for the electron's "Zitterbewegung", which leads to an effective "smearing" of the electron wavefunction. [47,72,73] The Darwin correction term is:

$$H_D = \frac{\hbar^2}{8{m_e}^2c^2}\nabla^2 V(\vec{r} - \vec{r}') \tag{65}$$

We should take a closer look on the expression $\vec{r} - \vec{r}'$. For the central Coulomb potential of the core, this leads to $\vec{r}' = 0$ and one obtains:

$$\nabla^2 V(\vec{r}) = 4\pi e^2\delta^3(\vec{r}) \quad \Leftrightarrow \quad \langle E_D\rangle = \frac{e^2}{\varepsilon_0}\frac{\hbar^2}{8{m_e}^2c^2}{f_N}^2|\alpha_0|^3 \tag{66}$$

Hence, the Darwin term only affects states with non-zero probability density at the core of the $\delta-$function, i.e., the center of the charge.

### d. *The Lamb shift*

The Lamb shift represents a quantum electrodynamic correction to atomic energy levels that cannot be explained within the framework of the Dirac equation alone. It arises from radiative corrections due to Self-energy and vacuum polarization. [29-35,46,71,72,73] In the helium atom, these corrections still apply individually to each electron, but they are now embedded in a correlated, interacting system. The dominant contribution to the Lamb shift for low-Z atoms like helium is the self-energy correction, which in leading logarithmic approximation takes the Hamiltonian form:

$$H_L(\vec{r}-\vec{r}') = \frac{1}{6\pi}\frac{e^2}{4\pi\varepsilon_0 \hbar c}\frac{\hbar^2}{{m_e}^2c^2}\left[\ln\left(\frac{m_e c^2}{E_{He}}\right) + \ln\left(\frac{1}{\alpha^2}\right) - \gamma - \frac{5}{6}\right]\nabla^2 V(\vec{r}-\vec{r}') \tag{67}$$

This results in the energy terms for the core analogue to (66)

$$\Leftrightarrow \langle E_L\rangle = \frac{Z\alpha^2}{6\pi}\frac{\hbar^3}{{m_e}^2 c}\left[\ln\left(\frac{m_e c^2}{E_{He}}\right) + \ln\left(\frac{1}{\alpha^2}\right) - \gamma - \frac{5}{6}\right]{f_N}^2|\alpha_0|^3 \tag{68}$$

### e. *Spin-Spin contributions due to the Breit Hamiltonian*

The relativistically consistent description of the interaction between electrons in many-particle systems requires correction terms derived from quantum electrodynamics. [41,42,75,76] The Breit operator, originally introduced by Gregory Breit, incorporates relativistic corrections to the electron-electron interaction by using the exchange of a virtual photon in the Coulomb gauge and leads to an effective correction to the Dirac Hamiltonian. The spin-spin interaction is a direct consequence of the magnetic field produced by the spin of one electron acting on the magnetic moment of the other. In quantum electrodynamics it appears as a low-order relativistic correction to the instantaneous electron-electron interaction. In our case only the Fermi contact term is relevant.:

$$H_{S-S} = \frac{2}{3}\mu_0 {g_e}^2 {\mu_B}^2 \boldsymbol{s_i s_j}\delta^3(|r_{12}|) = -\frac{\mu_0 {g_e}^2 e^2\hbar^2}{8\pi {m_e}^2}\delta^3\big(Re(z) - Im(z)\big) \tag{69}$$

With the spin operators $\boldsymbol{s_i}, \boldsymbol{s_j}$ being $\boldsymbol{s_i s_j} = -\frac{3}{4}\hbar^2$ for the singlet state *S*=0. Also, dipole terms, as well as spin-orbit and orbit-orbit interactions vanish. If we consider the mutual interaction of two electrons, the term develops into the form: $\nabla^2 V(\vec{r}-\vec{r}') = \nabla^2 V\left(\frac{1}{|r_{12}|}\right)$. The expression $|r_{12}|$ can in our case be rewritten as $Re(z) - Im(z) = x - y$ as shown in (70) which gives a convenient way for further proceeding.

$$\nabla^2 V\left(\frac{1}{Re(z) - Im(z)}\right) = 4\pi e^2 \int_{\bar{\gamma}} \delta^3\big(Re(z) - Im(z)\big)dz \quad with\, \bar{\gamma}: x = y, \forall z \in {\mathbb{C}^+}_0$$
$$\Leftrightarrow \int_{\bar{\gamma}}\left(\int_{\gamma} \delta^3\big(Re(z) - Im(z)\big)\overline{R_c(z)}R_c(z)z^2 dz\right)dz = \int_0^\infty R_c(x)^2 dx = f_V \tag{70}$$

For $\forall z \in \mathbb{C}^{+}{}_{0}$ and reformulating $\nabla^2 V\left(\frac{1}{|r_{12}|}\right)$ utilizing (70) we collect all $\delta$-functions localized on the angle bisector of the complex plane. We choose the integration way based on the prove from chapter 9:

$$\Leftrightarrow \langle E_{S-S} \rangle = -\frac{\mu_0 {g_e}^2 e^2 \hbar^2}{8\pi {m_e}^2} f_N f_V |\alpha_0|^3 \tag{71}$$

(71) can now be used very conveniently to calculate the perturbation energy of the spin-spin-interaction.

## 7. Energy determination of the neutral Helium

As already mentioned and referred to in previous chapters, we will take a deeper look into the energy calculation. The completely written-out Schrodinger equation then takes the following form:

$$\langle R_{He}|\hat{H}_c|R_{He}\rangle = E \tag{72}$$

Since only the S states with their spherical symmetry are considered, (72) is reduced to the radial part. This results in the following formula:

$$\langle R_c|\hat{H}_c|R_c\rangle = \int_\gamma \overline{R_c(z)}\frac{\partial^2}{\partial z^2}R_c(z)z^2dz + 2\int_\gamma z\overline{R_c(z)}\frac{\partial}{\partial z}R_c(z)dz + \int_\gamma z^2\overline{R_c(z)}\left(\frac{A}{z} - B|E|\right)R_c(z)dz - C_E\int_\gamma z^4R_c(z)^2\overline{R_c(z)}R_c(z)dz \tag{73}$$

For the full documentation please refer to the MATLAB script shown in chapter 12. As $R_c$ contains the decay rate $\alpha_0$ and thus the energy expression intrinsically referring to (73), the Schrodinger equation cannot be mathematically resolved according to $\alpha_0$ and needs to be solved iteratively.

The fundamental ground energy is calculated separately, but for a complete picture also the energy perturbation terms discussed in chapter 6, i.e. the Darwin term, Lamb shift and spin-spin interaction as well as the proposed vibrational ground mode are considered in the iteration. All derivatives of $R_c(z)$ can be calculated analytically, though the expressions are somewhat unwieldy. For convenience, we use the MATLAB function diff() from the Symbolic Toolbox. The integrals, on the other hand, are solved numerically by vpaintegral() in MATLAB, as there is no analytical master function for some integrals.

By inserting the literature ground state energy straight ahead as a seed into the equation, we already receive a decent result, meaning we can reproduce the value to an accuracy of -1.0007 eV, but to prove that the solution converges, the MATLAB file in chapter 12 solves (73) in 47 iterations until the energy deviation drops below a defined threshold of $10^{-9}$ (relative error between iterations, respectively). Below this threshold numerical noise caused instabilities limiting further precision. In our view, though, there is no principal obstacle to achieve arbitrary precision levels for the calculation with more advanced computer capabilities. Unfortunately, they were not available to the author when writing this article. The script produces within 47 iterations a ground state energy purely derived by ab initio conditions of

$$\langle \boldsymbol{E_0}\rangle = \boldsymbol{-79.00830183374404519040837018720244239 4(62)\ eV} \tag{74}$$

That is a relative deviation from the literature value of:

$$\frac{\boldsymbol{\Delta E}}{\langle \boldsymbol{E_{lit}} \rangle} = \frac{\langle \boldsymbol{E_0} \rangle - \langle \boldsymbol{E_{lit}} \rangle}{\langle \boldsymbol{E_{lit}} \rangle} \approx -\mathbf{0.0000235370838881693453} \cong -\mathbf{23.5}\, \boldsymbol{ppm} \tag{75}$$

And an absolute error of:

$$\boldsymbol{\Delta E} = \langle \boldsymbol{E_0} \rangle - \langle \boldsymbol{E_{lit}} \rangle \approx \mathbf{1.859668799}\, \boldsymbol{meV} \tag{76}$$

So, the theoretical approach presented here can reproduces the lowest S state. One can arise the question as to why only the ground state of helium was evaluated, while the hydrogen solution was able to derive the entire energy spectrum with all quantum states. The answer is close at hand: The S states are exotic in the sense that two electrons interact to the maximum extent and shows correspondingly the highest repulsive value – therefore the strongest effect. Moreover, due to this phenomenon, the S states are not orthogonal to mixed excited states, which in consequence makes a closed formulation of the energy spectra much more difficult. An excited state of the helium atom such as $He^*(1s^0 2p^1)$ has a significantly lower overlap – and thus interaction - of the two electrons, so a lower coupling factor would appear. Such a state can be described with acceptable accuracy by the superposition of two hydrogen states - as described in the literature [10,14-17]. The smaller the overlap of the two electron wave functions is, the better is the approximation. This report thus closes to a certain extent the gap between the already well calculable excitation and ionization states and the various approximations via variational calculations from the literature for the ground state.

## 8. Conclusions

In this report an analytical solution of the Schrödinger equation for neutral helium in the ground state $^1S_0$ was presented. Neither perturbation calculus nor variational methods are used, but the differential equation is solved directly by analytical means using the Laplace transformation and the result is evaluated.

The fundamental properties of the wave function for both electrons were investigated, and it was shown that it is possible to formulate an electron state function, which is not a classical Hartree mean field approach, but covers both electrons in their entangled state in a novel formulation using complex analysis. A general formula of an electron potential term in the form of a radial probability density was derived by utilizing Green's formalism and convolving Coulomb's potential with the correlated electron density function.

Additionally, we integrated the energy contribution of a dark electronic vibrational mode as well as contributions of the spin-spin interaction originating from Breit terms and the Lamb shift evaluated for the helium system by perturbation calculations.

Although already consistent in simple forward calculation, we verified the results by iteration of the wave function in the Schrodinger equation, thus proving a consistent and stable ground state with a deviation of $\Delta E \approx +1.859668799\ meV$ referring to the literature value:

The analytical solution of the Schrödinger equation for helium generates a wave function similar, but not identical to the ground state of hydrogen. In particular, the electron probability density is much more compacted at the nucleus, both because the maximal probability radius lies at ${a_0}^{He} = 1.58219417134664422(78) * 10^{-11}\ m$ and the spatial decay rate is higher than hydrogen by a factor of $\alpha_0 a_0 = 2.40977241078937160(43)$.

We proposed a dark vibrational electronic mode as a significant energy contributor to the ground energy and – even more important - as an important energy absorption mechanism which could help explaining chemical inertness of helium and the phenomenon of the closed shell.

In our view, the mathematical methods presented in this work open a wide range of possibilities, whereby this report only makes a small contribution. Since geometrically structures are almost always found finite in nature, it seems promising to investigate further mathematical and physical problems that are currently infested with singularities and check whether the method presented here could possibly offer a suitable solution.

## 9. Appendix: Transforming $\boldsymbol{R_{He}(\vec{r}_1,\vec{r}_2) \to R_c(z)}$ and performing integration

To transfer the original helium system consisting of two electrons, thus consequently formulated in 6 dimensions, we choose the following approach: Transfer the two spatial vectors of both electrons in one complex vector considering all spatial information:

$$\vec{r}_z = \vec{r}_1 + i\vec{r}_2$$

So, this results in:

$$\vec{r}_z = \begin{pmatrix} r_1 + ir_2 \\ \varphi_1 + i\varphi_2 \\ \theta_1 + i\theta_2 \end{pmatrix}$$

Considering, that the wavefunction is spherical symmetric and transferring it to the complex function $R_{He}(\vec{r}_1,\vec{r}_2) \to R_c(z)$, this means that $R_c$ does not depend on $\varphi_i, \theta_i$ and for the radial part we can define:

$$z = x + iy, \qquad x \stackrel{\text{def}}{=} r_1 \quad and \quad y \stackrel{\text{def}}{=} -ir_2 \quad with \quad r_1, r_2 \in \mathbb{R}^+$$

Using the premises from chapter 3, the wavefunction must be symmetric

$$R_{He}(r_1,r_2) = R_{He}(r_2,r_1)$$

and holomorphic

$$\frac{\partial}{\partial z}R_c(z) = \frac{1}{2}\left(\frac{\partial R_c}{\partial x} - i\frac{\partial R_c}{\partial y}\right) = \frac{1}{2}\left(\frac{\partial R_c}{\partial x} - i^2\frac{\partial R_c}{\partial x}\right) = \frac{\partial R_c}{\partial x} = -i\frac{\partial R_c}{\partial y}$$

The last claim is plausible and motivated by the fact that the impulse operator must be well defined and finite for every direction on the definition range. Claiming holomorphy has far-reaching consequences:

- $R_c$ has a well-defined complex derivative at every point in its domain.
- It is infinitely differentiable (smooth) in the complex sense.
- It can be locally represented by a convergent power series.
- The integral over any closed contour in a simply connected domain is zero and values inside a contour can be calculated using values on the contour.

So holomorphy is an even stronger demand than total differentiability for real functions. Additionally, $R_c$ is also square integrable and declines sufficient fast for $z \to \infty$. The next step is now evaluating the expectation value of an operator to solve the Schrodinger equation:

$$\langle R_{He}(\vec{r}_1,\vec{r}_2)|O|R_{He}(\vec{r}_1,\vec{r}_2)\rangle = \langle R_c(z)|O|R_c(z)\rangle = \int_\gamma \overline{R_c(z)}OR_c(z)d^3z$$

Where $\overline{R_c(z)}$ is the complex conjugated of $R_c(z)$

We transfer the differential $d^3\vec{r}_1 d^3\vec{r}_2 \to d^3\vec{r}_z$ and move to spherical coordinates: $d^3\vec{r}_z = 4\pi z^2 dz$. So, the system is transferred from a two-particle problem with 6 coordinates to a formal one particle problem and reformulated in spherical coordinates. The next step is switching over to a complex path integral by evaluating $\overline{R_c(z)}\mathcal{O}R_c(z)$ and choosing a proper path $\gamma$:

Be the path $\gamma = \gamma_1 + \gamma_2 + \gamma_3$ defined in $\mathbb{C}_0{}^+$ with $a \in \mathbb{R}_0{}^+$:

$$\gamma_1: y\,runs\,from\;0 \to a\;, \qquad x = 0$$

$$\gamma_2: x\,runs\,from\;0 \to b \to \infty, \qquad y = a$$

$$\gamma_3: y\,runs\,from\;a \to \infty, \qquad x = b \to \infty$$

As said before, $R_c(z)$ declines sufficiently quick for vast x and y to 0. So, the result gives

$$\int_\gamma z^2 \overline{R_c(z)}\mathcal{O}R_c(z)dz$$

$$= i\int_0^a (0+iy)^2\overline{R_c(0-\imath y)}\mathcal{O}R_c(0+iy)dy$$

$$+\int_0^\infty (x+ia)^2\overline{R_c(x+\imath a)}\mathcal{O}R_c(x+ia)dx$$

$$+\,i\int_a^\infty (b+iy)^2\overline{R_c(b-\imath y)}\mathcal{O}R_c(b+iy)dy$$

Note, that $ia$ is not the complex conjugate in the argument, but the imaginary constant coordinate of the path.

$$\Leftrightarrow \langle R_c|\mathcal{O}|R_c\rangle = i\int_0^a (iy)^2\overline{R_c(-\imath y)}\mathcal{O}R_c(iy)dy + \int_0^\infty (x+ia)^2\overline{R_c(x+\imath a)}\mathcal{O}R_c(x+ia)dx$$
$$+\,i\int_a^\infty (b+iy)^2\overline{R_c(b-\imath y)}\mathcal{O}R_c(b+iy)dy$$

We substitute now $y = -iy'and\;dy = -idy'$. Substitute analogue $a \to -ia$, i.e. we use the symmetry and perform a Wick-rotation. Remember, that the argument of $R_c$ turns real by that and complex conjugation is now irrelevant.

$$\Leftrightarrow \langle R_c|\mathcal{O}|R_c\rangle = \int_0^a y'^2 R_c(-y')\mathcal{O}R_c(y')dy' + \int_0^\infty (x+a)^2R_c(x+a)\mathcal{O}R_c(x+a)dx$$
$$+\int_a^\infty (b+y')^2R_c(b-y')\mathcal{O}R_c(b+y')dy'$$

Now we implement the substitution $x' = x + a$ and fit to the integration limits.

$$\Leftrightarrow \langle R_c|O|R_c\rangle = \int_0^a y'^2 R_c(-y')OR_c(y')dy' + \int_a^\infty x'^2 R_c(x')OR_c(x')dx$$
$$+ \int_a^\infty (b+y')^2 R_c(b-y')OR_c(b+y')dy'$$

Analyzing $R_c(-y')$ in detail as the actual function that resulted from the Schrodinger equation in chapter 5, we find for x=0, y variable:

$$R_c(-y') = f_N \frac{sinh(-\alpha_0 y')}{(-\alpha_0 y')\left(1 + AShi(-\alpha_0 y')\right)^2} = +f_N \frac{sinh(\alpha_0 y')}{\alpha_0 y'\left(1 + AShi(\alpha_0 y')\right)^2} = R_c(y')$$

As $sinh(-\alpha_0 y') = -sinh(\alpha_0 y')$ and

$$Shi(-\alpha_0 y') = \int_0^{y'} \frac{sinh(-\alpha_0 \xi)}{-\alpha_0 \xi} d\xi = \int_0^{y'} \frac{-sinh(\alpha_0 \xi)}{-\alpha_0 \xi} d\xi = Shi(\alpha_0 y')$$

So, as a consequence, the following statement must be true:

$$\Leftrightarrow \langle R_c|O|R_c\rangle = \int_0^a y'^2 R_c(y')OR_c(y')dy' + \int_a^\infty x'^2 R_c(x')OR_c(x')dx$$
$$+ \int_a^\infty (b+y')^2 R_c(b-y')OR_c(b+y')dy'$$

We now take the limit of the third integral as *b* approaches infinity:

$$\lim_{b\to\infty} \int_a^\infty (b+y')^2 R_c(b-y')OR_c(b+y')dy' = 0$$

So, this results in:

$$\Leftrightarrow \langle R_c|O|R_c\rangle = \int_0^a y'^2 R_c(y')OR_c(y')dy' + \int_a^\infty x'^2 R_c(x')OR_c(x')dx \equiv \int_0^\infty x^2 R_c(x)OR_c(x)dx$$

As a>0 can be chosen freely, $\langle R_c|O|R_c\rangle$ can be completely evaluated by the right integral, independent of the chosen path in the complex plane. Though $\overline{R_c(-\imath y)}OR_c(iy)$ is not holomorphic, yet the integral is independent on the chosen path for $R_c$ resulting in a quasi-holomorphic behavior. With this proven, we write down every integral like this:

$$\langle R_c|R_c\rangle = \int_0^\infty R_c(x)^2 x^2 dx \tag{77}$$

$$\int_\gamma \overline{R_c(z)} \frac{\partial^2}{\partial z^2} R_c(z) z^2 dz = \int_0^\infty R_c(x) \left[\frac{\partial^2}{\partial x^2} R_c(x)\right] x^2 dx \tag{78}$$

$$\int_{\gamma} \frac{2}{z}\overline{R_c(z)}\frac{\partial}{\partial z}R_c(z)z^2dz = 2\int_0^{\infty} xR_c(x)\left[\frac{\partial}{\partial x}R_c(x)\right]dx \tag{79}$$

$$\int_{\gamma} z^2\overline{R_c(z)}\left(\frac{A}{z} - B|E|\right)R_c(z)dz = \int_0^{\infty} x^2\left(\frac{A}{x} - B|E|\right)R_c(x)^2dx \tag{80}$$

$$C_E\int_{\gamma} z^4R_c(z)^2\overline{R_c(z)}R_c(z)dz = C_E\int_0^{\infty} x^4R_c(x)^4dx \tag{81}$$

## 10. Appendix: Expanding $R_c$ to a polynomial

Chapter 4 referred to the fact, that every potential caused by electrons present in the vicinity of the atom can be modelled as a sum over $n$ of a polynomial expression multiplied with an exponential damped probability depending on the distance to the core with the specific parameters $A_n$ and $a_n$. It is now to prove, that this is as well the case for the derived wave function (50) in chapter 6. Now if one uses reformulation of the equation, it is easy to find that:

$$R_c(z) = f_N \frac{sinh(\alpha_0 z)}{\alpha_0 z\left(1 + AShi(\alpha_0 z)\right)^2} = f_N \frac{e^{\alpha_0 z} - e^{-\alpha_0 z}}{2\alpha_0 z\left(1 + AShi(\alpha_0 z)\right)^2} \tag{82}$$

$$= f_N \left[\frac{e^{\alpha_0 z}}{2\alpha_0 z\left(1 + AShi(\alpha_0 z)\right)^2} - \frac{e^{-\alpha_0 z}}{2\alpha_0 z\left(1 + AShi(\alpha_0 z)\right)^2}\right]$$

$$= f_N [R_c'(z) e^{\alpha_0 z} - R'_c(z) e^{-\alpha_0 z}] \tag{83}$$

$$with \;\; R_c'(z) \equiv \frac{1}{2\alpha_0 z\left(1 + AShi(\alpha_0 z)\right)^2}$$

As $R_c(z)$ is holomorphic and therefore $R'_c(z)$ as well, they can be expanded locally into a Taylor power series as every holomorphic function. This power series converges against the function in a neighborhood of the development point. In addition, the Taylor expansion of a holomorphic function is unique. Moreover, $R_c(z)$ and $R'_c(z)$ are globally analytic on the domain of definition.

So, it is obvious to see that (83) is equivalent to expression (21) and that the approach to implement $R_c(z)$ into the Schrödinger equation is valid. Hence it is also proven, that (50) is a valid solution of the Schrödinger equation. qed.

## 11. Appendix: Holomorphy and symmetry of $R_c$

a) First, we must prove that $R_c(z)$ is holomorphic. So, using (50) we obtain:

$$R_c(z) = f_N \frac{sinh(\alpha_0 z)}{\alpha_0 z\left(1 + AShi(\alpha_0 z)\right)^2} = f_N \frac{sinh\left(\alpha_0(x+iy)\right)}{\alpha_0(x+iy)\left(1 + AShi\left(\alpha_0(x+iy)\right)\right)^2}$$

We use the Cauchy Riemann equations in form of the Wirtinger derivative for the proof [13,14]. So, it must be valid:

$$\frac{\partial}{\partial \bar{z}} R_c(z) = \left(\frac{\partial}{\partial x} + i\frac{\partial}{\partial y}\right) R_c(z) = 0 = f_N \frac{\partial}{\partial \bar{z}}\left[\frac{sinh\left(\alpha_0(x+iy)\right)}{\alpha_0(x+iy)\left(1 + AShi\left(\alpha_0(x+iy)\right)\right)^2}\right]$$

We write down the partial derivatives:

$$\frac{\partial}{\partial x} R_c(z) = AShi\left(\alpha_0(x+iy)\right)\frac{\left(\alpha_0(x+iy)cosh\left(\alpha_0(x+iy)\right) - sinh\left(\alpha_0(x+iy)\right)\right)}{\alpha_0(x+iy)^2\left(1 + AShi\left(\alpha_0(x+iy)\right)\right)^3}$$
$$+ \frac{\alpha_0(x+iy)cosh\left(\alpha_0(x+iy)\right)}{\alpha_0(x+iy)^2\left(1 + AShi\left(\alpha_0(x+iy)\right)\right)^3}$$
$$- \frac{sinh\left(\alpha_0(x+iy)\right)\left(1 + 2Asinh\left(\alpha_0(x+iy)\right)\right)}{\alpha_0(x+iy)^2\left(1 + AShi\left(\alpha_0(x+iy)\right)\right)^3}$$

And

$$\frac{\partial}{\partial y} R_c(z) = iAShi\left(\alpha_0(x+iy)\right)\frac{\left(\alpha_0(x+iy)cosh\left(\alpha_0(x+iy)\right) - sinh\left(\alpha_0(x+iy)\right)\right)}{\alpha_0(x+iy)^2\left(1 + AShi\left(\alpha_0(x+iy)\right)\right)^3}$$
$$+ i\frac{\alpha_0(x+iy)cosh\left(\alpha_0(x+iy)\right)}{\alpha_0(x+iy)^2\left(1 + AShi\left(\alpha_0(x+iy)\right)\right)^3}$$
$$- i\frac{sinh\left(\alpha_0(x+iy)\right)\left(1 + 2Asinh\left(\alpha_0(x+iy)\right)\right)}{\alpha_0(x+iy)^2\left(1 + AShi\left(\alpha_0(x+iy)\right)\right)^3}$$

Then obviously $\frac{\partial}{\partial \bar{z}} R_c(z) = \frac{\partial}{\partial x} R_c(z) + i\frac{\partial}{\partial y} R_c(z) = 0$. So $R_c(z)$ is holomorphic. qed.

b) Secondly, we must prove that $R_c(z)$ is symmetric according to the coordinates x and y in the complex plane, considering the substitution that we switch real and imaginary part. This means that $R_c(x+iy) = R_c(y+ix)$.

We start with the following equation utilizing the trigonometric transformations with $\alpha > 0$:

$$\frac{sinh\big(\alpha_0(x+iy)\big)}{\alpha_0(x+iy)} = \frac{cos(\alpha_0 y)sinh(\alpha_0 x) + isin(\alpha_0 y)cosh(\alpha_0 x)}{\alpha_0(x+iy)} \tag{84}$$

We now transform the coordinates between real and imaginary axis on the right side according to

$$x \longrightarrow ix, \qquad iy \longrightarrow y \Leftrightarrow y \longrightarrow -iy$$

With the trigonometric relations:

$$sinh(ia) = isin(a), \qquad cosh(ia) = cos(a), \qquad sin(ia) = isinh(a), \qquad cos(ia) = cosh(a)$$

This results in:

$$\Leftrightarrow \frac{sinh\big(\alpha_0(x+iy)\big)}{\alpha_0(x+iy)} = \frac{cos(-i\alpha_0 y)sinh(i\alpha_0 x) + isin(-i\alpha_0 y)cosh(i\alpha_0 x)}{\alpha_0(x+iy)}$$

$$\Leftrightarrow \frac{sinh\big(\alpha_0(x+iy)\big)}{\alpha_0(x+iy)} = \frac{icosh(\alpha_0 y)sin(\alpha_0 x) + iisinh(-\alpha_0 y)cos(\alpha_0 x)}{\alpha_0(x+iy)}$$

$$\Leftrightarrow \frac{sinh\big(\alpha_0(x+iy)\big)}{\alpha_0(x+iy)} = \frac{cos(\alpha_0 x)sinh(\alpha_0 y) + isin(\alpha_0 x)cosh(\alpha_0 y)}{\alpha_0(x+iy)} = \frac{sinh\big(\alpha_0(y+ix)\big)}{\alpha_0(x+iy)} \tag{85}$$

So obviously $\frac{sinh(\alpha_0 z)}{\alpha_0 z}$ is symmetric under coordinate transformation. But then also $Shi(\alpha_0 z)$ is symmetric as well. Thus, the necessary conclusion is $R_c(x+iy) = R_c(y+ix)$. qed.

## 12. Appendix: MATLAB-Code for evaluation of the Schrodinger equation

The MATLAB script below uses the standard functions and the symbolic toolbox as well to calculate derivatives and integrals of the helium wavefunction. After defining the physical constants and the substitutions to simplify the calculations, the while loop iteratively calculates the energy and uses the value as the next iteration point for alpha, thus evaluation the wave function and the Hamiltonian, until the energy deviation of two iterations drops below the threshold. The literature value of helium is used as iteration starting point and the results are also compared with this value afterwards. Note, that $R_{He}(z)$ is substituted to $R_{He}(\mathfrak{z})$ with $\mathfrak{z} = \frac{z}{\alpha_0}$ to simplify the numerical integration. That is also the reason for the prefactor $\alpha_0$ in the energy calculations without changing the results. It also results in $AShi(\alpha_0 z) \rightarrow \frac{A}{\alpha_0} Shi(z')$ due to substitutions in the integral term. Finally, all energy contributions are listed for comparison and transparency. The results are converted into a string to make arbitrary number of digits available.

```
clear
close all
clc
tic;
digits(200);  % define precision

syms x alpha_sym;

% physical constants (as vpa)
hquer               = vpa(6.62607015e-34/(2*pi));                          % Heisenbergs constant
C                   = vpa(299792458);                                       % speed of light
e_e                 = vpa(1.602176634e-19);                                 % elementary charge
m_e                 = vpa(9.1093837015e-31);                                % electron mass
Epsilon0            = vpa(8.8541878128e-12);                                % dielectric constant
Z                   = vpa(2);                                               % charge of helium
E_H                 = vpa(Z^2 * m_e * e_e^4 / (2*(4*pi*Epsilon0)^2*hquer^2)); % Hydrogen ground state energy
E_He_lit            = vpa(-24.587389011 * e_e - E_H);                       % Helium ground state energy
from literature
g_e                 = vpa(2.00231930436092);                                % g-factor of the electron
alpha_S             = vpa(e_e^2/(4*pi*Epsilon0*hquer*C));                   % Sommmerfeld fine structure
constant
a0                  = 4 * pi * Epsilon0 * hquer^2 / (e_e^2 * m_e);          % Bohr radius
% calculation constants (with vpa precision)
B                   = 2 * m_e / hquer^2;
A                   = B * Z * e_e^2 / (4 * pi * Epsilon0);
C_C                 = A * 4;

% iteration parameters
threshold           = vpa(1e-9);                                            % abort criteria for iteration
x_min               = 0;
x_max               = 1;
E_test              = E_He_lit;
E_result            = E_test;
E_diff              = vpa(1);
i                   = 0;

% Symbolic definition of Rc
R_sym             = sinh(x)/(x * (1 + A / alpha_sym * sinhint(x))^2);

% Iteration loop with vpa precision
while abs(E_diff) > threshold
    E_test          = E_result;
    alpha           = vpa(sqrt(- B * E_test));%vpa(sqrt(C_C - B * E_test));                         % cal-
culation of alpha
    % Symbolic calculation of derivatives
    R_sub           = subs(R_sym, alpha_sym, alpha);
    dR_sub          = diff(R_sub, x);
    ddR_sub         = diff(dR_sub, x);
    dddR_sub        = diff(ddR_sub, x);
    ddddR_sub       = diff(dddR_sub, x);
    % electron potential
    phi_ee          = x^2 * R_sub^2;
    % spherical laplace operator of the electron potential
    ddphi_ee_sub  = 2 * R_sub^2 + 2 * x^2 * (dR_sub^2 + R_sub * ddR_sub);
```

```
    % Relativistic impulse correction with Bi-Laplace
    Bi_LP_KK_sub  = alpha^4 * ddddR_sub + 2 * alpha^4 / x * dddR_sub + 2 * alpha^3 / x * ddR_sub;
    % Symbolic integration
    % Normation
    f_N_square           = vpa(1 / vpaintegral(x^2 * R_sub^2, x, 0, Inf));
    % Maximum of the wave-function-square
    a_He            = vpa(vpasolve(diff(x^2 * R_sub^2) == 0, x, [x_min x_max]));
    % Impulse terms
    E_kin_1         = vpa(alpha^2 * vpaintegral(x^2 * R_sub * ddR_sub, x, 0, Inf));
    E_kin_2         = vpa(2 * alpha^2 * vpaintegral(x * R_sub * dR_sub, x, 0, Inf));
    E_rel           = hquer^2/(4 * m_e^2 * C^2) * vpa(vpaintegral(x^2 * R_sub * Bi_LP_KK_sub, x, 0, Inf)); % Relativ-
istic Impulse correction
    % Kinetic energy
    E_kin           = - (E_kin_1 + E_kin_2 + E_rel);
    % Coulomb term of the core
    E_C             = vpa(- A * alpha * vpaintegral(x * R_sub^2, x, 0, Inf));
    % Coulomb term of the elektrons
    E_ee            = vpa(f_N_square * C_C * alpha * vpaintegral(x^2 * R_sub^2 * phi_ee, x, 0, Inf));
    % Zero point energy of the elctronic vibration mode
    k_vib_C         = vpa(diff(x * R_sub^2));
    k_vib_ee        = vpa(diff(x^4 * R_sub^4));
    k_a_He_C        = vpa(-A * f_N_square / alpha * subs(k_vib_C, x, double(a_He)));
    k_a_He_ee       = vpa(- C_C * f_N_square^2 * subs(k_vib_ee, x, double(a_He)));
    k_a_He          = k_a_He_C + k_a_He_ee;
    E_vib           = vpa(hquer * sqrt(k_a_He/(m_e)) / 2);
    % pertubational energy terms
    E_result_base = vpa(- f_N_square * (E_kin + E_C + E_ee ) / B);
    % Darwin term
    E_Dar           = vpa(hquer^2 * e_e^2 / (8 * m_e^2*C^2 * Epsilon0) * f_N_square * alpha^3);
    % Lambshift
    E_Lamb          = vpa(alpha_S^2 * 4 * pi * hquer^3/ (6 * pi * 8 * m_e^2 * C) * f_N_square * alpha^3  * (log(-
(m_e*C^2)/E_He_lit) - log(Z^2*alpha_S^2) - eulergamma - 5/6));
    % Spin-Spin interaction
    f_V             = vpa(vpaintegral(R_sub^2, x, 0, Inf));
    E_SS            = - vpa(g_e^2 * e_e^2 * hquer^2 /  ( 8 * pi * Epsilon0 * m_e^2  * C^2) * f_V * f_N_square * alpha^3
);
    % result
    E_result        = E_result_base + 2 * E_vib + E_Dar + E_Lamb + E_SS;
    % Iteration and calculation of deviation related to abort criterium
    E_diff          = (E_result - E_test)/E_result;
    double(E_diff)
    i               = i + 1; % # of loops
end
% output (with full precision as text)
fprintf('Iterations: %d\n', i);
fprintf('Energy: %s eV\n', char(vpa(E_result / e_e, 40)));
fprintf('rel. tolerance (absolute): %s (%s ppm)\n', char(vpa(abs((E_result - E_He_lit)/E_He_lit), 20)),
char(vpa(abs((E_result - E_He_lit)/E_He_lit)*1e6, 3)));
fprintf('abs. tolerance (absolute): %s meV\n\n', char(vpa(abs((E_result - E_He_lit)/e_e*1000), 10)));
fprintf('spherical electron radius: %s pm\n\n', char(vpa(a_He / alpha *1e12, 20)));
fprintf('spatial decay rate alpha: %s m^-1\n\n', char(vpa(alpha, 20)));
fprintf('Difference decay rates He / H: %s \n\n', char(vpa(a0*alpha, 20)));

t = vpa((E_result_base/e_e),4);
fprintf('contribution ground energy:         %s eV\n',char(t));
t = vpa((2 * E_vib/e_e),4);
fprintf('contribution vibrational zero energy: %s eV\n',char(t));
t = vpa((E_Dar/e_e),4);
fprintf('contribution Darwin-term:             %s eV\n',char(t));
t = vpa((E_Lamb/e_e),4);
fprintf('contribution Lamb-Shift:              %s eV\n',char(t));
t = vpa((E_SS/e_e),4);
fprintf('contribution Spin-Spin-interaction:  %s eV\n',char(t));
toc
```

## 13. Glossary

The following list shows the important mathematical functions, parameters and constants in alphabetical order:

| **Symbol** | **Meaning** |
|---|---|
| $\alpha$ | Fine structure constant |
| $\alpha_0$ | Decay rate of the radial helium wavefunction |
| $a_0$ | Bohr radius |
| ${a_0}^{He}$ | maximum of the electron probability density / covalent radius for helium |
| $A$ | Substitution as $\frac{2m_e}{\hbar^2}\frac{Ze^2}{4\pi\varepsilon_0}$ |
| B | Substitution as $\frac{2m_e}{\hbar^2}$ |
| c | Speed of light |
| $C_e$ | Substitution as $\frac{2m_e}{\hbar^2}4\frac{Ze^2}{4\pi\varepsilon_0}$ |
| $\delta$ | Distance of two interacting electrons $\delta = \vert r_1 - r_2\vert$ |
| $e$ | Elementary charge |
| $\varepsilon_0$ | Electrical field constant |
| $E_n$ | Energy of Helium from the Hamiltonian in $n^{th}$ state |
| $E_D$ | Energy of Darwin perturbation term |
| $E_L$ | Energy of Lamb shift perturbation term |
| $E_{S-S}$ | Energy of spin-spin perturbation term |
| $f_{N,n}$ | normalization factor of the radial helium wavefunction |
| $\hat{H}_1, \hat{H}_2$ | Separated Hamiltonians for both electrons |
| $\hat{H}_c$ | Adapted Hamiltonian with transformed electron interchange potential |
| $\hat{H}_{ee}$ | Hylleraas Hamiltonian with $r_{12}$ term |
| ${\hat{H}_{ee}}^{z}$ | transformed electron interchange potential written in complex parameters |
| $\hbar$ | Reduced Planck constant |
| $\lambda_C$ | Compton wavelength |
| $L$ | Overall angular momentum of both electrons |
| $m_e$ | Electron rest mass |
| $k_B$ | Boltzmann's constant |
| $\vert\phi\rangle\vert\psi\rangle$ | General Product wavefunction of two separate wavefunctions |
| $\vert\phi,\psi\rangle$ | General entangled wavefunction |
| $\phi(\vec{r})$ | Coulomb potential of a point like source |
| $R_{He}(r_1, r_2)$ | Radial helium wavefunction with real distances of the electrons $r_1, r_2$ |
| $R_c(z)$ | Radial helium wavefunction written with complex parameter $z = x + iy$ |
| $\Re_c(s)$ | Laplace transform of the radial helium wavefunction |
| ${}^1S_0$ | Neutral helium ground state |
| S | Overall spin of both electrons |
| $V(z)$ | Radial symmetric electron potential energy, written in complex terms |
| $V_E(z)$ | Radial symmetric electron potential energy with coupling factor |

## 14. Statements and Declarations

No funding was received to assist with the preparation of this manuscript. The author has no relevant financial or non-financial interests to disclose.
All data referred to in this work are available or accessible via the sources in the bibliography. Data referred to via the code can be accessed by executing the given code in chapter 12.
The Author of this work is fully accountable for all parts of this work including the code, ensuring that questions related to the accuracy or integrity of any part of the work are appropriately investigated and resolved. In this work, generative AI is used solely to make editorial and grammatical improvements. The content (text, figures and tables) was created entirely by the author of this work.